\documentclass[twocolumn,showpacs,aps,prb,letterpaper,nofootinbib,raggedbottom,nobalancelastpage]{revtex4}
\usepackage{graphicx,bm}
\usepackage{amsmath}

\usepackage{color}

\begin{document}
\title{Effect of phonon coupling on the generated entangled states of photons from a single quantum dot embedded inside a microcavity}
\date{\today}
\author{J. K. Verma, Harmanpreet Singh and P. K. Pathak}
\address{School of Basic Sciences,
Indian Institute of Technology Mandi, Kamand, H.P. 175005, India}
\begin{abstract}
We discuss generation of two types of entangled state of two photons-- noon state which is entangled in number and polarization, and polarization entangled state which is entangled in polarization and frequency. We consider a single quantum dot coupled with a bimodal cavity in strong coupling regime. We analyze the effect of exciton-phonon coupling  on concurrence of the generated entangled states. We find that for both states concurrence is maximum in the absence of anisotropic energy gap between exciton states and remains unchanged in the presence of exciton-phonon coupling. However for finite anisotropic energy gap concurrence decreases on increasing temperature of phonon bath.
\end{abstract}
\maketitle

\section{Introduction}
A source of entangled photons is an essential building block of quantum technologies such as  quantum metrology\cite{metrology1,metrology2}, quantum lithography\cite{lithography1,lithography2}, quantum imaging\cite{qimaging} and quantum information processing\cite{qinf}. The photons employed in these experiments are entangled in two degrees of freedom and are undistinguishable otherwise, for example entangled state in number and polarization such as noon state $|\psi_{noon}\rangle=\frac{1}{\sqrt2}(|n_x,0_y\rangle+|0_x,n_y\rangle)$, where the subscripts $x$, $y$ are corresponding to two orthogonal polarizations and $n$, $0$ represent number of photons in the corresponding mode.  Noon states have been found particularly successful for achieving uncertainty limited high resolution in quantum metrology and quantum lithography\cite{metrology1,metrology2}. Another most widely used state is entangled in polarization and frequency, called polarization entangled state, such as $|\psi_{pol}\rangle=\frac{1}{\sqrt2}(|x_1,x_2\rangle+|y_1,y_2\rangle)$, where subscript $1$, $2$ are corresponding to two frequencies $\omega_1$, $\omega_2$ and $x$, $y$ are two orthogonal polarizations.
In most of the experiments to date the noon states and polarization entangled states are generated by the parametric down conversion(PDC)\cite{qimaging,pdc2,pdc3}. The rate of generation of entangled photons from such sources is very low and probabilistic. However, in quantum technologies\cite{sqinf} a scalable on demand source of entangled photons is desirable. Recently, the quantum dots have been recognized as a potential candidate for realizing solid state sources of single photons\cite{singlephoton} and polarization entangled photons \cite{photopair1,photopair2,photopair3,photopair4}. In quantum dots, the cascaded decay of biexciton state to ground state via two exciton states having angular momenta $\pm1$, generate the polarization entangled photon state. However, the anisotropic energy gap between two exciton states in quantum dots breaks the indistinguishability between emitted x-polarized and y-polarized photon pairs as a result the
entanglement between generated photon pairs is washed out. Therefore, for
on chip realization of entangled photons using QD, it is essential to minimize the anisotropic energy gap. Several  techniques have been employed to minimize the anisotropic energy gap in QDs, for example, by applying external electric field\cite{electricfield1,electricfield2} or magnetic field\cite{magneticfield, elecmag}, by applying a proper combination of stress  and electric field\cite{elecstress1,elecstress2,elecstress3}, and using thermal annealing\cite{annealing}. In order to enhance the collection efficiency it is desirable to couple QD with cavity. However, due to large biexciton binding energy, it is difficult to couple biexciton to exciton and exciton to ground state transitions with the same cavity mode resonantly. There have been some interesting experiments to minimize or even making binding energy positive by using lateral electric field and by controlled fabrication\cite{binding1, binding2}.

As the QDs are solid state materials, the interaction of excitons with phonons in the QD-cavity system is inevitable. The exciton-phonon interaction leads to dephasing processes\cite{reservoir1,reservoir2,reservoir3,reservoir4} which
can destroy the entanglement between the generated photon pairs. It has also been found that interaction with phonons can enhance far off-resonant transitions, leading to cavity mode feeding. In recent years, various experiments have been performed in QD-cavity systems using incoherent\cite{photo1,photo2,photo3} and coherent excitation\cite{resonant1,resonant2,resonant3} and recording photoluminescence spectra. It has been observed that in coherent excitation, contrary to incoherent excitations\cite{photo1}, various dephasing phenomena such as coupling with multi-excitons and charged excitons do not influence the exciton-cavity dynamics and only the effects of exciton-phonon coupling are significant. The exciton-phonon coupling mainly leads to pure dephasing and off-resonant cavity mode coupling.  However, it has been shown that pure dephasing does not affect entanglement seriously as long as 'which path information' does not get imprinted on phonon bath\cite{reservoir1}. The effect of pure dephasing can also be minimized by increasing the cavity decay rate under strong coupling between QD and cavity. On the other hand, a significantly different nature of QD-cavity coupling has been observed in Mollow triplets from off resonant QDs due to phonon induced off-resonant cavity coupling. Recently, Majumdar et al.\cite{resonant4} have also observed that phonon induced off-resonant cavity coupling could be strong enough to transfer  excitation between a resonantly driven QD and a off-resonant QD placed inside the same cavity. In this paper, we discuss two different conditions for generating two photon noon state and polarization entangled state using QD-cavity system and analyze the effect of exciton-phonon coupling on concurrence of the generated states. We extend recently developed master equation formalism\cite{meq}, which has been successfully used in analyzing data of recent experiments.
\section{Theory}
\label{Sec:Theory}
\begin{figure}[h!]
\centering
\includegraphics[width=8cm]{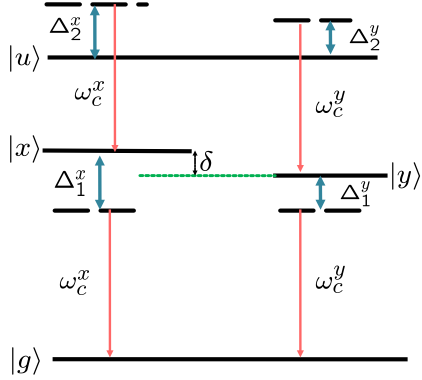}
\vspace{-0.1cm}
\caption{(Color online)
The energy level diagram of coupled quantum dot-cavity system for generating two photon noon state and polarization entangled state. The biexciton state $|u\rangle$ to  exciton state $|x\rangle$ ($|y\rangle$) and exciton state $|x\rangle$ ($|y\rangle$) to
ground state $|g\rangle$ transitions are coupled with x-polarized (y-polarized) cavity mode.}
\label{fig1}
\end{figure}
We consider a single quantum dot embedded in a photonic crystal microcavity.  The quantum dot consist of a ground state $|g\rangle$, the exciton energy levels $|x\rangle$, $|y\rangle$ and a biexciton energy level $|u\rangle$ as shown in Fig.1.
The anisotropic energy gap between the exciton energy levels is given by $\delta=\omega_x-\omega_y$, where $\omega_x$ and $\omega_y$ are the frequencies corresponding to the exciton states $|x\rangle$ and $|y\rangle$, respectively. The cavity has two orthogonally polarized modes of frequencies $\omega_{c}^x$ and $\omega_{c}^y$, respectively. The x-polarized mode is coupled to the $|u\rangle$ to $|x\rangle$ transition with coupling constant $g_{2}^x$ and the $|x\rangle$ to $|g\rangle$ transition with coupling constant $g_1^{x}$. Similarly the y-polarized mode is coupled to the $|u\rangle$ to $|y\rangle$ transition with coupling constant $g_{2}^y$ and the $|y\rangle$ to $|g\rangle$ transition with coupling constant $g_1^{y}$. The hamiltonian of the system in rotating frame with frequency $\omega_0=(\omega_x+\omega_y)/2$, is given by,
\begin{eqnarray}
H= -\hbar\Delta_{xx}|u\rangle\langle u|+\hbar\frac{\delta}{2}(|x\rangle\langle x|-|y\rangle\langle y|)\nonumber\\
	-\hbar\left(\Delta_1^x-\frac{\delta}{2}\right)a_x^\dag a_x
-\hbar\left(\Delta_1^y+\frac{\delta}{2}\right)a_y^{\dag}a_y\nonumber\\
+\hbar g_1^x(|x\rangle\langle g|a_x+a_x^\dag|g\rangle\langle x|)+\hbar g_2^x(|u\rangle\langle x|a_x+a_x^\dag|u\rangle\langle x|)\nonumber\\
	+\hbar g_1^y(|y\rangle\langle g|a_y+a_y^\dag|g\rangle\langle y|)+\hbar g_2^y(|u\rangle\langle y|a_y+a_y^\dag|y\rangle\langle u|)\nonumber\\
+H_{QD-Ph},~~
	\label{ham}
\end{eqnarray}
Where $\hbar\Delta_{xx}=\hbar(\omega_x+\omega_y-\omega_u)$ is the biexciton binding energy corresponding to biexciton frequency $\omega_u$, $\Delta_1^x=\omega_x-\omega_c^x$, $\Delta_2^x=\omega_{u}-\omega_x-\omega_c^x$ are the detunings from the x-polarized cavity mode, for exciton $|x\rangle$ to ground state $|g\rangle$ transition, the biexciton $|u\rangle$ to exciton $|x\rangle$ transition, and  $\Delta_1^y=\omega_y-\omega_c^y$, $\Delta_2^y=\omega_{u}-\omega_y-\omega_c^y$ are the detunings from the y-polarized cavity mode for exciton $|y\rangle$ to ground state $|g\rangle$ , the biexciton $|u\rangle$ to exciton $|y\rangle$ transition, respectively. The cavity field operators are given by $a_x$, $a_y$. The interaction of phonon bath with QD is given by
\begin{eqnarray}
H_{QD-ph}=\hbar\sum_k\omega_kb_k^{\dag}b_k+\hbar\sum_{\alpha=x,y,u}|\alpha\rangle\langle\alpha|\sum_k\lambda_k^{\alpha}(b_k^{\dag}+b_k),
\label{qdph}
\end{eqnarray}
where $b_k$ ($b_k^{\dag}$) are annihilation (creation) operator for k-th phonon mode of frequency $\omega_k$, and $\lambda_k^{\alpha}$ is the coupling of k-th phonon mode with exciton state $|\alpha\rangle$ for $\alpha=x,y,u$. For typical self assembled QDs the deformation potential coupling to longitudinal acoustic phonon is only significant. The coupling with transverse acoustic phonons and optical phonons do not couple significantly in InGaAs/GaAs self assembled QDs considered in this work\cite{stateprep}.
The QDs have large biexciton binding energy, therefore same cavity mode can not couple both the biexciton to exciton and exciton to ground state transitions resonantly. It has been noticed that for far off-resonantly coupled QD-cavity systems phonon interaction plays significant role and the phenomenon such as off-resonant cavity mode feeding and dephasing significantly alter the dynamics of the system. Therefore one can expect significant effect of phonon coupling on entangled states generated through biexciton decay. We are interested in coupled QD-cavity system which is described by reduced density matrix after tracing over phonon modes. The dynamics of the reduced density matrix is given by master equation derived using polaron transformed Hamiltonian to include phonon interaction of all order. We perform polaron transform $H'=e^SHe^{-S}$, where $S=\sum_{\alpha=x,y,u}|\alpha\rangle\langle\alpha|\sum_k\lambda_k^{\alpha}(b_k^{\dag}-b_k)$. The polaron transformed Hamiltonian can be rearranged as $H'=H_s+H_b+H_{sb}$, with
\begin{eqnarray}
H_s=-\hbar\Delta_{xx}|u\rangle\langle u|+\hbar\frac{\delta}{2}(|x\rangle\langle x|-|y\rangle\langle y|)\nonumber\\
	-\hbar\left(\Delta_1^x-\frac{\delta}{2}\right)a_x^\dag a_x -\hbar\left(\Delta_1^y+\frac{\delta}{2}\right)a_y^{\dag}a_y+\hbar\langle B\rangle X_g\\
H_b=\hbar\sum_k\omega_kb_k^\dag b_k\\
H_{sb}=\zeta_gX_g+\zeta_uX_u,
\label{poltransform}
\end{eqnarray}
where $\langle B\rangle=\langle B_-\rangle=\langle B_+\rangle$ with phonon displacement operator $B_\pm=\exp\left[\pm\sum_k\frac{\lambda_k}{\omega_k}(b_k-b_k^\dag)\right]$, $X_g=\sum_{\alpha=x,y}(g_1^\alpha|\alpha\rangle\langle g|a_{\alpha}+g_2^\alpha|u\rangle\langle\alpha|a_{\alpha})+H.c.$, $X_u=\sum_{\alpha=x,y}(ig_1^\alpha|\alpha\rangle\langle g|a_\alpha+ig_2^\alpha|u\rangle\langle\alpha|a_{\alpha})+H.c.$ are the QD-cavity interaction terms and $\zeta_g=(B_++B_-)/2-\langle B\rangle$, $\zeta_u=-i(B_+-B_-)/2$ are phonon fluctuation operators. The phonon induced frequency shifts in exciton and biexciton are absorbed in the detunings. The master equation  for reduced density matrix of QD-cavity coupled system is derived using Born-Markov approximation and phonon bath correlations in continuum limit using spectral density $J(\omega)=\alpha_p\omega^3\exp[-\frac{\omega^2}{2\omega_b^2}]$, where $\alpha_p$ is exciton phonon coupling and $\omega_b$ is the phonon cutoff frequency. In our simulations we use $\alpha_p=1.42\times10^{-3}g_1^{x2}$ and $\omega_b=10g_1^x$, which gives $\langle B\rangle=1.0,~0.90,~0.84$, and $0.73$ for $T=0K$, $5K$, $10K$, and $20K$, respectively. To include the spontaneous decay, cavity mode leakage and pure dephashing we add corresponding terms using Lindblad operator $L[\mu]\rho=\mu^\dag\mu\rho-2\mu\rho\mu^\dag+\rho\mu^\dag\mu$. The master equation for the coupled QD-cavity system is given by\cite{meq}
\begin{eqnarray}
\frac{\partial\rho}{\partial t}=-\frac{i}{\hbar}[H_s,\rho]-\frac{\kappa_x}{2}L[a_x]\rho-\frac{\kappa_y}{2}L[a_y]\rho-\frac{\gamma_1}{2}L[|g\rangle\langle x|]\rho\nonumber\\
-\frac{\gamma_1}{2}L[|g\rangle\langle y|]\rho-\frac{\gamma_2}{2}L[|x\rangle\langle u|]\rho-\frac{\gamma_2}{2}L[|y\rangle\langle u|]\rho\nonumber\\
-\frac{\gamma_d}{2}L[|x\rangle\langle x|]\rho-\frac{\gamma_d}{2}L[|y\rangle\langle y|]\rho-\gamma_dL[|u\rangle\langle u|]\rho\nonumber\\
-\frac{1}{\hbar^2}\int_0^\infty d\tau\sum_{m=g,u}\left(G_m(\tau)[X_m,e^{-iH_s\tau/\hbar}X_me^{iH_s\tau/\hbar}\rho]+H.c.\right)
\label{mseq}
\end{eqnarray}
where $\gamma_1$ and  $2\gamma_2$  are the spontaneous decay rate of exciton and biexciton states, $\kappa_x$ and $\kappa_y$ are the decay rate of x and y polarized cavity modes respectively, $\gamma_d$ is the pure dephasing rate. The phonon bath correlations are given by $G_g(\tau)=\langle B\rangle^2[\cosh\phi(\tau)-1]$ and $G_u(\tau)=\langle B\rangle^2\sinh\phi(\tau)$ with  $\phi(\tau)=\int_0^{\infty}d\omega \frac{J(\omega)}{\omega^2}[\coth(\frac{\hbar\omega}{2K_BT})\cos\omega t-i\sin\omega t]$.
The dynamics of the system is simulated by integrating master equation (\ref{mseq}) using quantum optics toolbox\cite{toolbox}. Here we mention that there have been other exciton-phonon coupling models which have been used particularly for externally driven QDs. For example correlation expansion approximation method\cite{correxp} and exact method using path integral approach\cite{pathint} have been developed. However, for the parameters used in this paper all such methods lead to same results. Therefore using convolutionless polaron master equation for numerical simulations in this work is well justified.

Recently Gl\"{a}ssl et al\cite{stateprep} have shown that for realistic parameters high fidelity robust generation of biexciton state preparation is possible due to phonon assisted off-resonant transition. In their proposal they applied a linearly polarized pulse of duration $15$ps applied resonantly between exciton and ground state which makes biexciton to exciton transition detuned by the biexciton binding energy. They have found that for biexciton binding energy range $0.5$Mev - $2.0$Mev and sufficiently large pulse areas almost perfect biexciton state generation with probability close to $1$ is possible. Further the process is facilitated by exciton-phonon coupling the generation becomes faster on increasing temperature. In the following sections we consider initially QD is prepared in biexciton state which decays to ground state by emitting two photons through two orthogonally polarized cavity modes.

\section{Generation of two photon NOON State}
 For generating two-photon noon state, $|\psi_{noon}\rangle\sim\frac{1}{\sqrt{2}}(|2_x,0_y\rangle+|0_x,2_y\rangle)$, which is entangled in number and polarization degrees of freedom, we choose frequencies of cavity modes $\omega_c^x=\omega_c^y=\omega_u/2$. The x-polarized cavity mode satisfies the two photon resonance condition, $\Delta_1^x+\Delta_2^x=0$  and similarly the y-polarized cavity mode satisfies the two photon resonance condition $\Delta_1^y+\Delta_2^y=0$. Because of large biexciton binding energy in QDs, $\left|\Delta_1^x\right|>>g_1^x, \left|\Delta_2^x\right|>>g_2^x$ and $\left|\Delta_1^y\right|>>g_1^y,
\left|\Delta_2^y\right|>>g_2^y$, therefore single photon transitions are significantly inhibited. Further, we consider that the interaction between QD and cavity modes is in strong coupling regime i.e. $(\kappa_x\ll g_{1}^x,g_{2}^x,~\kappa_y\ll g_{1}^y,g_{2}^y)$. Under these conditions the biexciton in QD decays via two paths by creating a photon pair either in x-polarized cavity mode or in y-polarized cavity mode. In first decay path, the state  $|u,0_x,0_y\rangle$ decays to  $|g,2_x,0_y\rangle$ via  $|x,1_x,0_y\rangle$ and in second decay path $|u,0_x,0_y\rangle$ decays to $|g,0_x,2_y\rangle$ via  $|y,0_x,1_y\rangle$. In the presence of leakage from cavity modes, the intermediate states $|x,1_x,0_y\rangle$ and $|y,0_x,1_y\rangle$ decay to $|x,0_x,0_y\rangle$ and $|y,0_x,0_y\rangle$ through single photon emission. Similarly, due to spontaneous decay, biexciton state $|u,0_x,0_y\rangle$ decays to states $|x,0_x,0_y\rangle$ or $|y,0_x,0_y\rangle$, thus reducing the two-photon transitions and enhancing single photon transitions. Therefore, it is desirable to work under strong coupling regime.  Using master equation (\ref{mseq}), we simulate the evolution of the system. We consider initial state of the coupled QD-cavity system as $|u,0_x,0_y\rangle$, i.e. QD is in biexciton state and there are no photons in the cavity modes.
\begin{figure}[h!]
\centering
\includegraphics[width=8cm, height=8cm ]{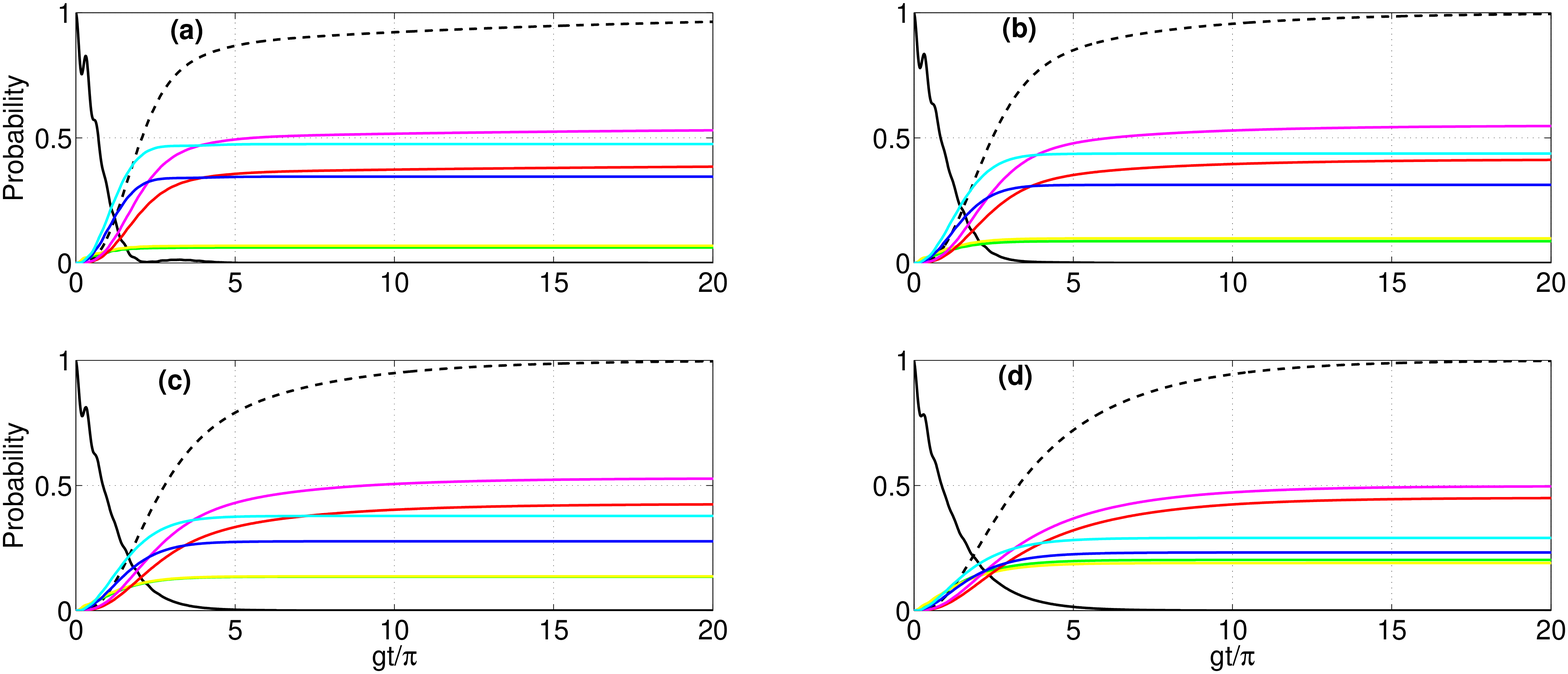}
\vspace{-0.1cm}
\caption{(Color online) The population in state $|u,0_x,0_y\rangle$ (black solid line), the population in state $|g,0_x,0_y\rangle$ (black dashed line), the emission probability $P_{1}^x$(green line) from state $|x,1_x,0_y\rangle$, the emission probability $P_{1}^y$ (yellow line) from state $|y,0_x,1_y\rangle$, the emission probability $P_{2}^x$ (red line) from state $|g,1_x,0_y\rangle$, the emission probability $P_{2}^y$ (magenta line) from the state $|g,0_x,1_y\rangle$, the emission probability $P_{3}^x$ (blue line) from the state $|g,2_x,0_y\rangle$ and the emission probability $P_{3}^y$(cyan line) from the state $|g,0_x,2_y\rangle$. The simulations are performed using parameters $g_1^x=g_2^x=g_1^y=g_2^y$, biexciton binding energy $\Delta_{xx}=10g_1^x$, anisotropic energy gap $\delta=g_1^x$, $\Delta_1^x=(\Delta_{xx}+\delta)/2$, $\Delta_1^y=(\Delta_{xx}-\delta)/2$, $\kappa_x=\kappa_y=0.4g_1^x$, $\gamma_1=\gamma_2=\gamma_d=0.01g_1^x$. The phonon bath temperature is for subplots (a) $T=0K$, (b) $T=5K$, (c) $T=10K$ and (d) $T=20K$.}
\label{fig2}
\end{figure}
In Fig.2, we plot population in the biexciton state $\langle u,0_x,0_y|\rho|u,0_x,0_y\rangle$, population in the ground state, $\langle g,0_x,0_y|\rho|g,0_x,0_y\rangle$ and the probabilities of single photon emission, $P_1^x(t)=\kappa_{x}\int_0^tdt'\langle x,1_x,0_y|\rho(t')|x,1_x,0_y\rangle$ from the state $|x,1_x,0_y\rangle$,
$P_1^y(t)=\kappa_{y}\int_0^tdt'\langle y,0_x,1_y|\rho(t')|y,0_x,1_y\rangle$ from the state $|y,0_x,1_y\rangle$,  $P_2^x(t)=\kappa_{x}\int_0^tdt'\langle g,1_x,0_y|\rho(t')|g,1_x,0_y\rangle$ from the state $|g,1_x,0_y\rangle$, $P_2^y(t)=\kappa_{y}\int_0^tdt'\langle g,0_x,1_y|\rho(t')|g,0_x,1_y\rangle$ from the state $|g,0_x,1_y\rangle$ and probabilities of single photon emission from two-photon states $P_3^x(t)=2\kappa_{x}\int_0^tdt'\langle g,2_x,0_y|\rho(t')|g,2_x,0_y\rangle$ from the state $|g,2_x,0_y\rangle$, $P_3^y(t)=2\kappa_{y}\int_0^tdt'\langle g,0_x,2_y|\rho(t')|g,0_x,2_y\rangle$ from the state $|g,0_x,2_y\rangle$. The detuning for x-polarized cavity mode $\Delta_1^x=-\Delta_2^x=\omega_x-\omega_u/2$ and detuning for y-polarized cavity mode $\Delta_1^y=-\Delta_2^y=\omega_y-\omega_u/2$, thus $\Delta_1^x>\Delta_1^y$ for positive $\delta_x$. The two-photon coupling for y-polarized transitions $2g_1^yg_2^y/\Delta_1^y$ is greater than two-photon coupling for x-polarized transition $2g_1^xg_2^x/\Delta_1^x$ \cite{laussy}. Therefore the single photon emission probability from two-photon state $P_3^y$ is larger than $P_3^x$, in long time limit. For small spontaneous decays the probability of biexciton decay through two-photon transition remains more than $0.9$. For subplots Fig.\ref{fig2} (a), (b), (c) and (d), we consider exciton coupling with phonon bath at temperature $0K$, $5K$, $10K$, and $20K$, respectively. When the temperature of phonon bath is raised phonon assisted off-resonant single photon transitions through biexciton cascaded decay are enhanced, and two-photon transitions are reduced. As a result the probabilities $P_1^x(t)$(green line) and $P_1^y(t)$(yellow line) increase and the probabilities $P_3^x(t)$(blue line) and $P_3^y(t)$(cyan line) decrease. The probability of photon emission from states $|g,1_x,0_y\rangle$ ($|g,0_x,1_y\rangle$), satisfy $P_2^x\approx P_1^x+P_3^x$ ($P_2^y\approx P_1^y+P_3^y$). The population in state $|g,1_x,0_y\rangle$ ($|g,0_x,1_y\rangle$) comes after one photon is leaked from state $|g,2_x,0_y\rangle$ ($|g,0_x,2_y\rangle$), and also when state $|x,0_x,0_y\rangle$ ($|y,0_x,0_y\rangle$) decays to $|g,1_x,0_y\rangle$ ($|g,0_x,1_y\rangle$) through single photon emission in cavity mode.

\begin{figure}[h!]
\centering
\includegraphics[width=10cm]{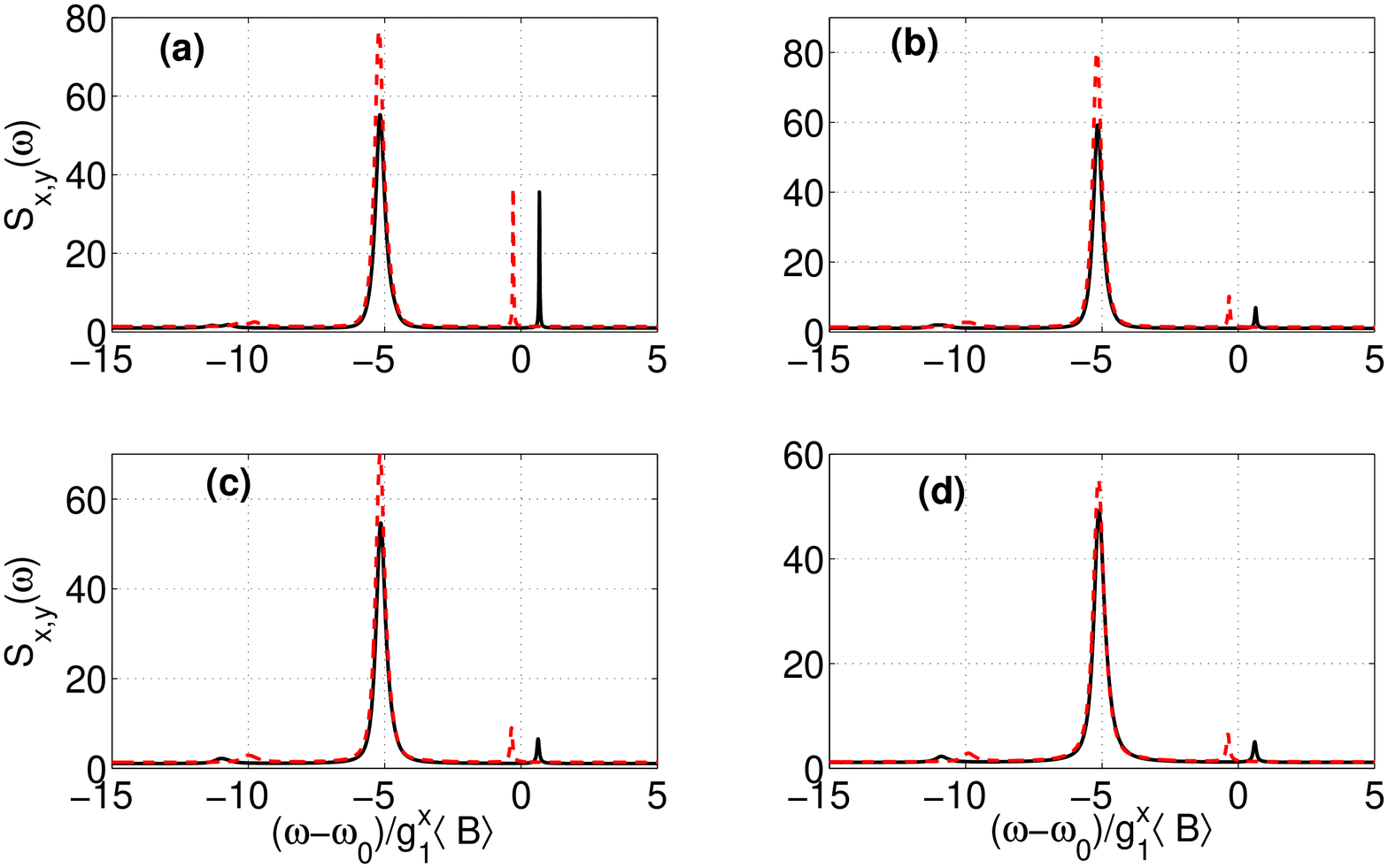}
\vspace{-0.1cm}
\caption{(Color online)
The spectrum of the photons emitted from x-polarized (black solid line) and y-polarized (red dashed line) cavity modes using same parameters as in Fig.\ref{fig2}. The frequencies of cavity modes are corresponding to $\omega_c^x-\omega_0=\omega_c^y-\omega_0=-\Delta_{xx}/2$.}
\label{fig3}
\end{figure}

We calculate the spectrum of emitted photons from the x and y polarized cavity modes which is given by,
\begin{equation}
    S_{x,y}(\omega)=\int_0^{\infty}dt\int_0^{\infty}d\tau \langle a_{x,y}^{\dagger}(t)a_{x,y}(t+\tau)\rangle e^{i\omega\tau}.
\label{sxy}
\end{equation}
The two time correlation $\langle a_{x,y}^\dag(t)a_{x,y}(t+\tau)\rangle$ is calculated using the quantum regression theorem. In Fig.\ref{fig3}, we present spectrum of photons emitted from cavity modes. The x-polarized and y-polarized photons show three peak spectrum. In the spectrum of x-polarized photons, exciton like peaks corresponding to single photon transition occur at $\omega-\omega_0\approx\delta/2$ and $\omega-\omega_0\approx-\Delta_{xx}-\delta/2$. Similarly in the spectrum of y-polarized photons there are exciton like peaks corresponding to $\omega-\omega_0\approx-\delta/2$ and $\omega-\omega_0\approx-\Delta_{xx}+\delta/2$. The central broad peaks in x-polarized and y-polarized at $\omega-\omega_0\approx-\Delta_{xx}/2$, are corresponding to emission from $|g,2_x,0_y\rangle$ and $|g,0_x,2_y\rangle$ generated from resonant two-photon transition. The two photon peaks in the spectrum of x-polarized and y-polarized photons overlap perfectly which makes emitted photons indistinguishable in frequency.
In subplots Fig.\ref{fig3} (a), (b), (c) and (d), when temperature is raised, the phonon assisted single photon transitions are enhanced\cite{prb-ent}. As a result the single photon transitions in biexciton-exciton cascaded decay also occur around frequencies of the cavity modes. Therefore emission around exciton-like peaks decreases and the emission around cavity mode frequencies increases progressively. It has been observed that for negative detuning the higher energy polariton state dominates in emission and for positive detuning lower energy polariton state dominates\cite{prx}. In our scheme for both x-polarized and y-polarized cavity modes, biexciton to exciton transitions are negatively detuned therefore the higher energy polariton states around cavity modes dominate in emission spectra. Similarly, exciton states to ground state transitions are positively detuned therefore the lower polariton states around cavity modes dominate.
\begin{figure}[h!]
\centering
\includegraphics[width=10cm]{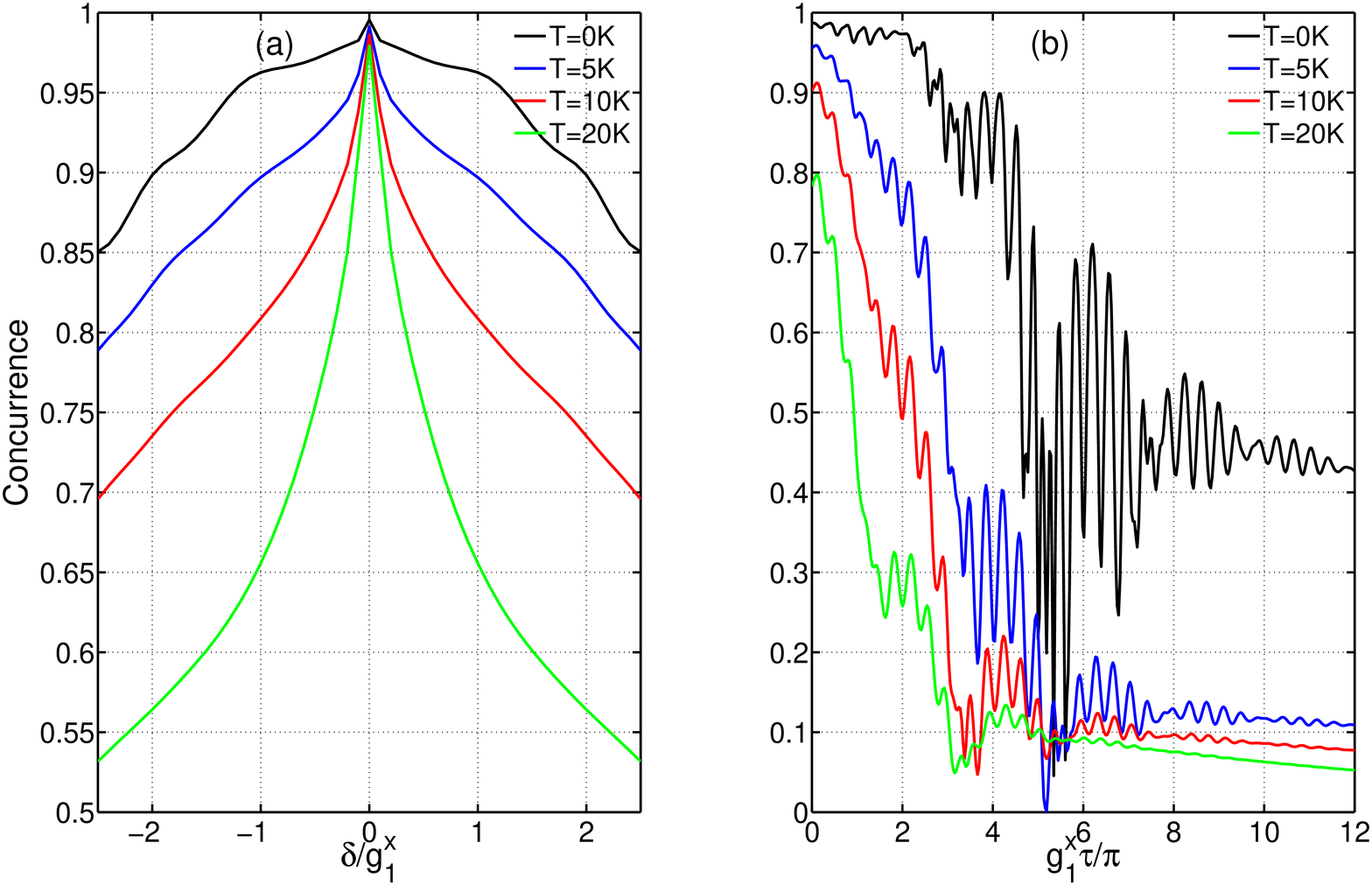}
\vspace{-0.1cm}
\caption{(Color online) In (a), the two time integrated concurrence $\bar{C}_{noon}$ defined as Eq.(\ref{cnc}) for the state of photons emitted through two-photon resonant transitions at different temperatures using parameters same as in Fig.\ref{fig2}. In (b), the single time integrated concurrence $C(\tau)$ defined by Eq.(\ref{ctau}).}
\label{fig4}
\end{figure}

In this case, as we discussed earlier initially the QD is prepared in the biexciton state which decays radiatively through resonant two-photon transitions via exciton states $|x\rangle$ or $|y\rangle$ and relaxes
in ground state $|g\rangle$ after emitting two photons in the entangled state,
\begin{equation}
    |\psi(t)\rangle_{noon}=\alpha_1(t)|2_x,0_y\rangle+\beta_1(t)|0_x,2_y\rangle,
\end{equation}
where $\alpha_1(t)$ and $\beta_1(t)$ are amplitudes corresponding to the generation of x-polarized and y-polarized photon pair. To reconstruct the density matrix of photon pair emitted from the cavity modes, coincidence measurements are performed. The photon coincidence measurements are given by two time correlation functions $G^2_{ij}(t,\tau)=\langle a^{\dag}_i(t)a^{\dag}_i(t+\tau)a_j(t+\tau)a_j(t)\rangle$, where $t$, $\tau$ are time of arrival of first photon at detector, delay time for second photon and $i$, $j$ are polarization of photons. Due to low collection, photons are detected for all arrival times and delay times. The reconstructed density matrix elements in polarization basis are given by $\rho_{iijj}\propto\int_0^{\infty}dt\int_0^{\infty}d\tau G^2_{ij}(t,\tau)$. The time independent concurrence for generated noon state can be defined as,
\begin{equation}
    \bar{C}=\frac{2\lvert \int_0^{\infty}d\tau \bar{G}_{xy}^2(\tau)\rvert}{\lvert\int_0^{\infty}d\tau \bar{G}_{xx}^2(\tau)\rvert+\lvert\int_0^{\infty}d\tau \bar{G}_{yy}^2(\tau)\rvert}
\label{cnc}
\end{equation}
Where,  $\bar{G}_{xy}^2(\tau)=\int_0^{\infty}\langle c_{1x}^{\dagger}(t)c_{2x}^{\dagger}(t+\tau)c_{2y}(t+\tau)c_{1y}(t)\rangle dt$,  $\bar{G}_{xx}^2(\tau)=\int_0^{\infty}\langle c_{1x}^{\dagger}(t)c_{2x}^{\dagger}(t+\tau)c_{2x}(t+\tau)c_{1x}(t)\rangle dt$
 and $\bar{G}_{yy}^2(\tau)=\int_0^{\infty}\langle c_{1y}^{\dagger}(t)c_{2y}^{\dagger}(t+\tau)c_{2y}(t+\tau)c_{1y}(t)\rangle dt$ are the second order two time correlation functions calculated with the help of quantum regression theorem. The operators $c_{i\alpha}$ for $i=1,2$ and $\alpha=x,y$ are defined as $c_{1x}=\sqrt{2}|g,1_x,0_y\rangle\langle g,2_x,0_y|$, $c_{2x}=|g,0_x,0_y\rangle\langle g,1_x,0_y|$, $c_{1y}=\sqrt{2}|g,0_x,1_y\rangle\langle g,0_x,2_y|$, and $c_{2y}=|g,0_x,0_y\rangle\langle g,0_x,1_y|$. In Fig.\ref{fig4}(a), we plot time independent concurrence with anisotropic energy gap between exciton states $\delta$ at different phonon bath temperatures. As we have found in Fig.\ref{fig2}, the presence of positive (negative) anisotropic energy gap leads to higher probability of y-polarized (x-polarized) two-photon transition which reduces concurrence slightly. Therefore in the absence of phonon bath interaction at $T=0K$, for parameters used in Fig.\ref{fig2} when single photon transitions are small, the maximum concurrence more than $0.95$ is achieved at $\delta=0$. On increasing $|\delta|$ the concurrence reduces slowly. However when phonon bath temperature is increased the phonon induced cavity mode feeding increases which enhances single photon transitions. Due to difference in frequencies of phonons absorbed during x-polarized and y-polarized photon pair emission, the which-path information gets imprinted on phonon bath which results larger reduction in concurrence. It has been shown in an earlier work that concurrence depending on photon arrival time $t$ given the more accurate value of entanglement between photons\cite{tconc}. However, in order to distinguish between photons generated through two-photon resonant process and through cascaded decay, one can detect photons for all arrival times $t$ but for different delay time $\tau$. The reconstructed such delay time dependent density matrix elements give time dependent concurrence\cite{tconc}.
 In Fig.\ref{fig4}(b), we plot time dependent concurrence defined in terms of cavity mode operators
 \begin{eqnarray}
 C(\tau)=\frac{2\lvert\int_0^{\infty}\langle a^{\dag}_x(t)a^{\dag}_x(t+\tau)a_y(t+\tau)a_y(t)\rangle dt\rvert}{\sum_{i=x,y}\int_0^{\infty}\langle a^{\dag}_i(t)a^{\dag}_i(t+\tau)a_i(t+\tau)a_i(t)\rangle dt}
 \label{ctau}
 \end{eqnarray}
 For small values of $\tau$, when photons are generated mainly from resonant two-photon transitions the concurrence remains large. However for larger values of $\tau$ when photons are generated through cascaded decay concurrence collapses to smaller values. For finite anisotropic energy gap $\delta$, x-polarized photon pair and y-polarized photon pair generated in cascaded decay become distinguishable which is visible even for $T=0$ when phonon interaction are neglected. We have checked that for $\delta=0$ such collapse does not appear in concurrence. At higher temperature when phonon assisted transitions take place the concurrence decreases, as a matter of fact, the phonons involved in x-polarized photon pair generation and y-polarized photon generation are distinguishable in frequencies for finite anisotropic energy gap thus revealing which path information. We notice that the effect of phonon interaction on concurrence has been discussed in earlier work\cite{heinze} where photons are generated through resonant two-photon process. However, they have considered weak coupling with the cavity mode so that probability for generating two photon state in cavity mode is negligible. It should be borne in mind that for entangled state of two photons photons should be distinguishable in two degrees of freedom.
\section{Generation of polarization entangled photon pair}
\begin{figure}[h!]
\centering
\includegraphics[width=10cm]{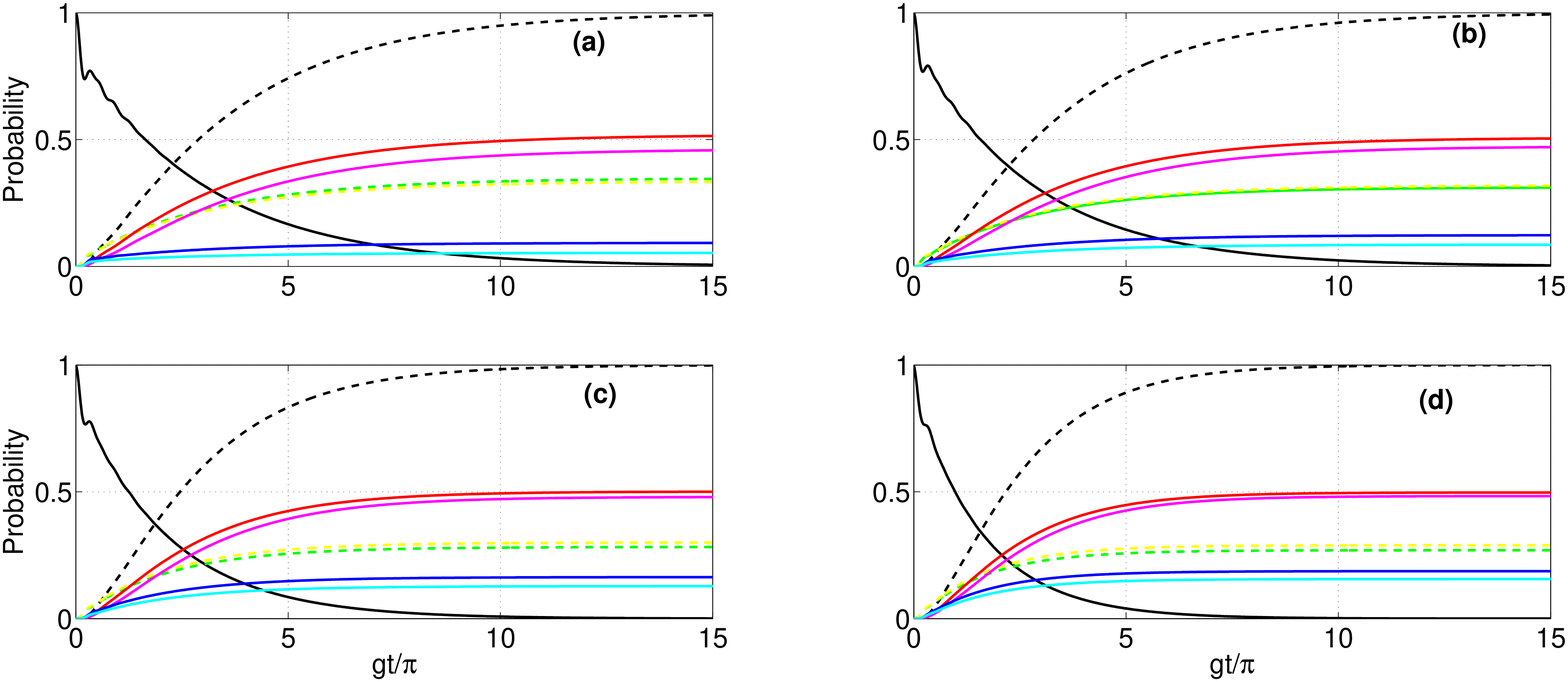}
\vspace{-0.1cm}
\caption{(Color online) The population in state $|u,0_x,0_y\rangle$ (black solid line), the population in state $|g,0_x,0_y\rangle$ (black dashed line), the emission probability $P_{1}^x$(green line) from state $|x,1_x,0_y\rangle$, the emission probability $P_{1}^y$ (yellow line) from state $|y,0_x,1_y\rangle$, the emission probability $P_{2}^x$ (red line) from state $|g,1_x,0_y\rangle$, the emission probability $P_{2}^y$ (magenta line) from the state $|g,0_x,1_y\rangle$, the emission probability $P_{3}^x$ (blue line) from the state $|g,2_x,0_y\rangle$ and the emission probability $P_{3}^y$(cyan line) from the state $|g,0_x,2_y\rangle$. The simulations are performed using parameters $g_1^x=g_2^x=g_1^y=g_2^y$, biexciton binding energy $\Delta_{xx}=5g_1^x$, anisotropic energy gap $\delta=g_1^x$, $\Delta_1^x=\delta$, $\Delta_1^y=-\delta$, $\kappa_x=\kappa_y=g_1^x$, $\gamma_1=\gamma_2=\gamma_d=0.01g_1^x$. The phonon bath temperature is for subplots (a) $T=0K$, (b) $T=5K$, (c) $T=10K$ and (d) $T=20K$.}
\label{fig5}
\end{figure}
For generating polarization entangled state $|\psi_{pol}\rangle\sim\frac{1}{\sqrt{2}}(|x_1,x_2\rangle+|y_1,y_2\rangle)$, we find the conditions when single photon transitions through cascaded decay will be dominating and the probability of two-photon transitions becomes negligible. The x-polarized photons are generated in the cascaded decay of biexciton by following single photon transitions $|u,0_x,0_y\rangle\rightarrow|x,1_x,0_y\rangle$ and $|x,0_x,0_y\rangle\rightarrow|g,1_x,0_y\rangle$. Similarly, the y-polarized photons are generated through $|u,0_x,0_y\rangle\rightarrow|y,0_x,1_y\rangle$ and $|y,0_x,0_y\rangle\rightarrow|g,0_x,1_y\rangle$ transitions. For maximum entanglement, the photons generated through biexciton to exciton transitions $|u,0_x,0_y\rangle\rightarrow|g,1_x,0_y\rangle$, $|u,0_x,0_y\rangle\rightarrow|y,0_x,1_y\rangle$ should have same frequency and the photons generated through excitons decay $|x,0_x,0_y\rangle\rightarrow|g,1_x,0_y\rangle$, $|y,0_x,0_y\rangle\rightarrow|g,0_x,1_y\rangle$ should also have same frequency. To match the frequencies we choose $\omega_c^x=\omega_y$ and $\omega_c^y=\omega_x$\cite{pathak1} so that $\Delta_1^x=-\Delta_1^y=\delta$ and $\Delta_2^x=\Delta_2^y=-\Delta_{xx}$. We also consider that QD has a moderate value of biexciton binding energy $\Delta_{xx}=5g_1^x$. The biexciton binding has been manipulated by using electric field and controlled fabrication\cite{binding1,binding2}.  In Fig.\ref{fig5}, we plot population in biexciton state $\langle u,0_x,0_y|\rho|u,0_x,0_y\rangle$, population in ground state $\langle g,0_x,0_y|\rho|g,0_x,0_y\rangle$ and the single photon extraction probabilities $P_1^x$, $P_1^y$, $P_2^x$, $P_2^y$, $P_3^x$, and $P_3^y$ as defined earlier. For typical parameters used in Fig.\ref{fig5}(a) in long time limit the value of $P_1^x=P_1^y\approx0.4$ which reflects that the probability of biexciton cascaded decay is $0.8$. Since biexciton to exciton transitions are highly detuned $\Delta_2^x=\Delta_2^y>4g_2^x,4g_2^y$ the transition rate for $|u,0_x,0_y\rangle\rightarrow|x,1_x,0_y\rangle$ and $|u,0_x,0_y\rangle\rightarrow|y,0_x,1_y\rangle$ are much smaller than cavity decay rates $\kappa_x,\kappa_y$. As a result most of the photons are leaked from states $|x,1_x,0_y\rangle$ and $|y,0_x,1_y\rangle$. The probability of generating two photon states $|g,2_x,0_y\rangle$ and $|g,0_x,2_y\rangle$ is very small. Therefore, the value of $P_2^x$ and $P_2^y$ remains smaller than $0.1$. Further the populations in $|g,1_x,0_y\rangle$ and $|g,0_x,1_y\rangle$ come from the decay of the states $|x,0_x,0_y\rangle$, $|g,2_x,0_y\rangle$ and $|y,0_x,1_y\rangle$, $|g,0_x,2_y\rangle$ respectively, that give $P_3^x\approx P_1^x+P_2^x$ and $P_3^y\approx P_1^y+P_2^y$. In subplots 5(b), 5(c) and 5(d), when the temperature of phonon bath is increased the phonon induced off-resonant biexciton to exciton transitions at cavity mode frequencies are enhanced leading to larger probabilities of generating two photon states $|g,2_x,0_y\rangle$ and $|g,0_x,2_y\rangle$. Therefore, at higher temperatures $P_1^x$, $P_1^y$ decrease and $P_2^x$, $P_2^y$ increase.
\begin{figure}[h!]
\centering
\includegraphics[width=10cm]{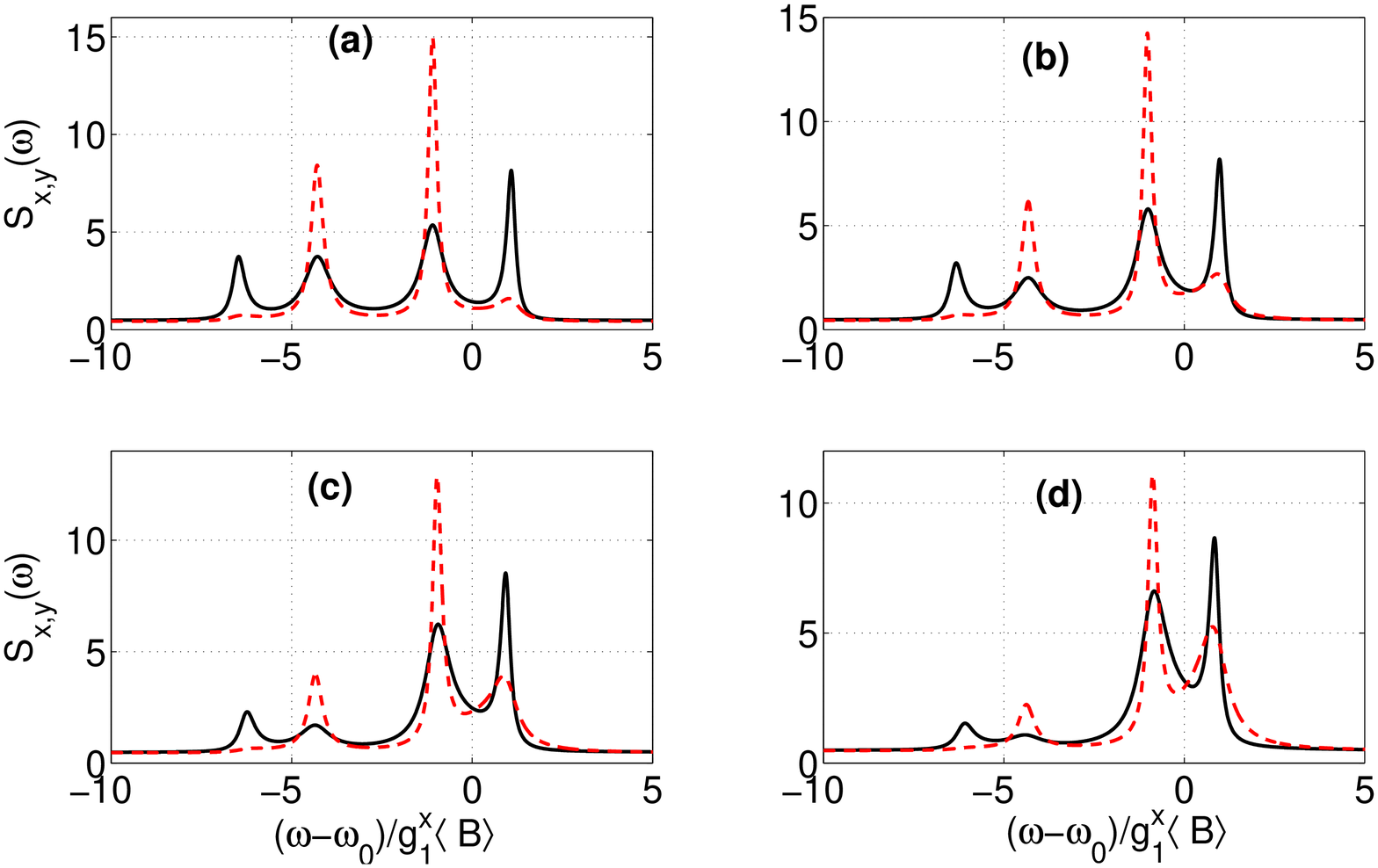}
\vspace{-0.1cm}
\caption{(Color online)The spectrum of the photons emitted from x-polarized (black solid line) and y-polarized (red dashed line) cavity modes using same parameters as in Fig.\ref{fig5}. The frequencies of cavity modes are $\omega_c^x-\omega_0=-\delta/2$, $\omega_c^y-\omega_0=\delta/2$}
\label{fig6}
\end{figure}

In Fig.\ref{fig6}, we plot spectrum of the emitted x-polarized and y-polarized photons calculated using Eq.(\ref{sxy}). The exciton states $|x\rangle$ and $|y\rangle$ are coupled with the cavity modes under strong coupling regime, as a result dressed polariton states with energies $\omega_x^\pm=-(\Delta_1^x-\delta\pm\sqrt{\Delta_1^{x2}+4g_1^{x2}})/2$ and $\omega_y^\pm=-(\Delta_1^y+\delta\pm\sqrt{\Delta_1^{y2}+4g_1^{y2}})/2$ are formed. The polariton states for x-polarized photons and for y-polarized photons become degenerate for $\Delta_1^x=-\Delta_1^y=\delta$. There are two peaks in the spectrum of x-polarized and y-polarized photons corresponding to two polariton dressed states at $\omega-\omega_0\approx\pm\sqrt{\delta^2+4g_1^{x2}}/2=\pm\sqrt{\delta^2+4g_1^{y2}}/2$. The biexciton to exciton transitions are highly detuned $g_2^x,g_2^y<\Delta_2^x,\Delta_2^y$. Therefore we treat coupling of biexciton state perturbatively\cite{pathak2}. The spectral peaks, corresponding to photons generated through biexciton state to polariton  state transitions, in x-polarized and y-polarized spectrum occur at $\omega-\omega_0\approx-(\Delta_{xx}+2g_2^{x2}/\Delta_{xx})-\omega_x^\pm$ and $\omega-\omega_0\approx-(\Delta_{xx}+2g_2^{y2}/\Delta_{xx})-\omega_y^\pm$ which are also degenerate. In subplots Fig.\ref{fig6} (b), (c) and (d), when temperature of the phonon bath is raised the probability for phonon assisted off-resonant transitions from biexciton state to exciton state (which is a linear superposition of polariton states) increases. The phonon assisted biexciton to exciton transitions occur around frequencies of cavity modes $\omega_c^x-\omega_0=-\delta/2$, $\omega_c^y-\omega_0=\delta/2$. Thus the emission around the frequencies $\omega-\omega_0\approx-(\Delta_{xx}+2g_2^{x2}/\Delta_{xx})-\omega_x^\pm$ and $\omega-\omega_0\approx-(\Delta_{xx}+2g_2^{y2}/\Delta_{xx})-\omega_y^\pm$ reduces and the emission around the cavity mode frequencies increases.
\begin{figure}[h!]
\centering
\includegraphics[width=10cm]{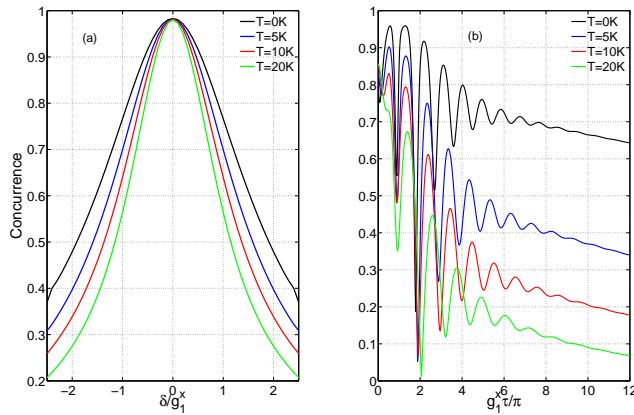}
\vspace{-0.1cm}
\caption{(Color online)In (a) time independent concurrence  and in (b) time dependent concurrence for polarization entangled state generated through single photon transitions at different temperatures, using parameters same as in Fig.\ref{fig5}.}
\label{fig7}
\end{figure}
In Fig.\ref{fig7}(a) we show time independent concurrence evaluated using Eq.(\ref{cnc})and $c_{1x}=|x,0_x,0_y\rangle\langle x,1_x,0_y|$, $c_{1y}=|y,0_x,0_y\rangle\langle y,0_x,1_y|$, $c_{2x}=|g,0_x,0_y\rangle\langle g,1_x,0_y|$, and $c_{2y}=|g,0_x,0_y\rangle\langle g,0_x,1_y|$. When there is no phonon interaction ($T=0$), the concurrence decreases on increasing anisotropic energy gap because the overlapping between the spectrum of x-polarized and y-polarized photons decreases\cite{pathak1,pathak2}. When the temperature of phonon bath is raised the phonon induced cavity mode feeding start dominating as a result more and more photons are generated through off-resonant biexciton to exciton transitions at cavity mode frequencies, ie. x-polarized photons are generated around frequency $\omega_c^x=\omega_y$ and y-polarized photons around frequency $\omega_c^y=\omega_x$.
For finite anisotropic energy gap these photons are distinguishable that gives which path information. Therefore for finite anisotropic energy gap there is more reduction in entanglement for the generated state due to phonon interaction. However, the x-polarized and y-polarized off-resonant biexciton to exciton transitions are equally detuned, therefore the phonons created in bath during x-polarized and y-polarized transitions can not provide which-path information. In Fig.\ref{fig7}(b), we show time dependent concurrence $C(\tau)$ calculated using Eq.(\ref{ctau}). In this case, the photon pair generated through two-photon transitions to the states $|g,0_x,2_y\rangle$, $|g,2_x,0_y\rangle$ and the photon pair generated through single photon cascaded decay are not differentiable in delay times due to cavity enhanced exciton decay. We notice that effect of phonon coupling on concurrence has been discussed by Heinze et al\cite{heinze} and Cygorek et al\cite{tconc} under strong coupling and weak coupling regimes. In their proposals two photons leaked from states $|g,0_x,2_y\rangle$, $|g,2_x,0_y\rangle$ and two photons generated through cascaded decay have been considered together with out difference as polarization entangled state.
\section{Conclusions}
\label{Sec:Conclusions}
We have discussed generation of two types of entangled state, which are entangled either in frequency and polarization or in number and polarization, in the system of QD embedded in a bimodal photonic crystal cavity. We have also discussed experimental conditions for generating different types of entangled state. The parameters used in our proposal could be for example $g_1^x=g_2^x=g_1^y=g_2^y=0.1meV$, $\kappa_x=\kappa_y=0.04meV-0.1meV$ and $\gamma_1=\gamma_2=\gamma_d=0.001meV$ which are feasible using recent technology\cite{photo2,photo3}. For temperature $T<10K$ and anisotropic energy gap $\delta$ less than cavity dipole coupling strength, concurrence larger than $0.7$, which is required for violation of Bell's inequalities\cite{bell}, can be achieved.  We have evaluated the effect of exciton-phonon coupling at different temperatures using recently developed master equation formulations. We have found that when anisotropic energy gap between exciton states is negligible coupling with phonon bath does not change entanglement significantly. However, for finite anisotropic energy gap which path information is revealed either by the frequencies of phonons or the frequencies of photons involved in phonon-assisted off-resonant transitions occurred during generation of photon pairs and the entanglement in the generated state is significantly reduced.
\section{Acknowledgements}
This work was supported by DST SERB Fast track young scientist scheme SR/FTP/PS-122/2011.

\end{document}